\documentclass{article}

\usepackage[utf8]{inputenc}
\usepackage{graphicx}
\usepackage[hyphens]{url}
\usepackage{framed}
\usepackage{amsmath}
\usepackage{tabularx}
\usepackage{listings}
\usepackage{color}
\usepackage{setspace}
\usepackage{xcolor}

\newcommand{\AAA}{AAA}
\newcommand{\AES}{AES}
\newcommand{\AP}{AP}
\newcommand{\APs}{APs}
\newcommand{\AS}{AS}
\newcommand{\ASCII}{ASCII}
\newcommand{\CA}{CA}
\newcommand{\CBC}{CBC}
\newcommand{\CNA}{CNA}
\newcommand{\CSR}{CSR}
\newcommand{\CSS}{CSS}
\newcommand{\EAP}{EAP}
\newcommand{\EAPoL}{EAPoL}
\newcommand{\EAPTLS}{EAP-TLS}
\newcommand{\EAPTTLS}{EAP-TTLS}
\newcommand{\EAPSH}{EAP-SH}
\newcommand{\FreeRADIUS}{FreeRADIUS}
\newcommand{\HMAC}{HMAC}
\newcommand{\HTML}{HTML}
\newcommand{\HTTP}{HTTP}
\newcommand{\HTTPS}{HTTPS}
\newcommand{\LAN}{LAN}
\newcommand{\IANA}{IANA}
\newcommand{\IETF}{IETF}
\newcommand{\IP}{IP}
\newcommand{\IPSec}{IPSec}
\newcommand{\ISPs}{ISPs}
\newcommand{\ISP}{ISP}
\newcommand{\IV}{IV}
\newcommand{\DNS}{DNS}
\newcommand{\DHCP}{DHCP}
\newcommand{\MAC}{MAC}
\newcommand{\MitM}{MitM}
\newcommand{\MSK}{MSK}
\newcommand{\NFC}{NFC}
\newcommand{\OAEP}{OAEP}
\newcommand{\OCSP}{OCSP}
\newcommand{\OSU}{OSU}
\newcommand{\OWE}{OWE}
\newcommand{\PKCS}{PKCS}
\newcommand{\PKG}{PKG}
\newcommand{\PKI}{PKI}
\newcommand{\PEAP}{PEAP}
\newcommand{\QR}{QR}
\newcommand{\RADIUS}{RADIUS}
\newcommand{\RFC}{RFC}
\newcommand{\RSA}{RSA}
\newcommand{\SHA}{SHA}
\newcommand{\SQLite}{SQLite}
\newcommand{\SSL}{SSL}
\newcommand{\TCP}{TCP}
\newcommand{\TLS}{TLS}
\newcommand{\UCA}{UCA}
\newcommand{\UAM}{UAM}
\newcommand{\URI}{URI}
\newcommand{\URL}{URL}
\newcommand{\WISPr}{WISPr}
\newcommand{\WIFI}{Wi-Fi}
\newcommand{\WLAN}{WLAN}
\newcommand{\WPAthree}{WPA3}
\newcommand{\XML}{XML}

\newcommand{\WPASupplicant}{wpa\_supplicant}
\newcommand{\openssl}{openssl}
\newcommand{\freeradius}{freeRADIUS}

\begin{document}

\title{Integration of the Captive Portal paradigm with the 802.1X architecture}
\author{
Nuno Marques,
André Zúquete,
João Paulo Barraca\\
\{nunommarques,andre.zuquete,jpbarraca\}@ua.pt}

\maketitle

\begin{abstract}

In a scenario where hotspot wireless networks are increasingly being
used, and given the amount of sensitive information exchanged on
Internet interactions, there is the need to implement security
mechanisms that guarantee data confidentiality and integrity in such
networks, as well as the authenticity of the hotspot providers.

However, many hotspots today use Captive Portals, which rely on authentication
through Web pages (thus, an application-level
authentication approach) instead of a link-layer approach. The
consequence of this is that there is no security in the
wireless link to the hotspot (it has to be provided at upper
protocol layers), and is cumbersome to manage wireless access profiles
(we need special applications or browsers' add-ons to do that).

This work exposes the weaknesses of the Captive Portals' paradigm,
which does not follow a unique nor standard approach, and describes
a solution that intends to suppress them, based on the 802.1X
architecture. This solution uses a new {\EAP}-compliant protocol
that is able to integrate an {\HTTP}-based registration or
authentication with a Captive Portal within the 802.1X
authentication framework.
\end{abstract}

\newcommand\keywords[1]{\textbf{Keywords}: #1}

\section{Introduction}

The security of 802.11 ({\WIFI}) wireless networks is defined by
several standards, such as 802.11i~\cite{802.11i} and 802.11w~\cite{802.11w}. These
standards define a wide range of security mechanisms, including
mutual authentication between a user and a network provider and
link-layer data encryption and frame integrity control. Furthermore, these
standards also open the path to evolution, namely to the development
of new, innovative authentication protocols. This is achieved with
the adoption of the 802.1X authentication architecture~\cite{802.1X}, which allows
different EAP (Extensible Authentication Protocol~\cite{rfc3748}) compliant
authentication protocols to be defined and deployed for tackling
special requirements~\cite{Dantu07}. A common characteristic of all these protocols
is that all of them produce a secret, shared session key that will be
used by 802.11i and 802.11w protection mechanisms to provide an
effective security at the link layer.  

One of the limitations of the actual, standard 802.11 security
paradigm is that it does not support the enrollment of new users by
a network. In fact, it supports the protection of users that already
share some access credentials with the network provider, but not the
setup of those credentials. This is a notorious limitation, since in
many cases users could benefit of using a {\WIFI} network that is
provided where they are, but they need first to enroll in that
network using some sort of alternative communication channel.

Captive Portals did the trick, and because of that, they became very
popular to deploy the so-called {\WIFI} hotspots. With Captive
Portals, the users may enroll on network providers using a Web
browser and the network of interest, and then use the same Captive
Portal to authenticate themselves prior to use the network. 
Recently Cisco forecast in a White Paper~\cite{Cisco2019} that
"\textit{the total public {\WIFI} hotspots (including homespots) will
grow four-fold from 2017 to 2022}", which means that the
exploitation of Captive Portals will continue to grow.

However, the exploitation of Captive Portals has several drawbacks. 
First, they are implemented today as an ad hoc, Web-based
Man-in-the-Middle ({\MitM}) hack that causes confusion in client
systems and users~\cite{Nomadix2013,httpwg2016,WirelessPhreak2017}.
Second, the Captive Portals' authentication is performed at
the applicational level, and thus does not produces key material for
being used by 802.11i and 802.11w. Therefore, users accessing
{\WIFI} networks controlled by Captive Portals are not protected at
the link layer. The consequence is that both users and 
access networks can be easily attacked and abused in many different ways,
which we will detail later (see Section~\ref{captive}). Therefore,
they are dangerous when it comes to the transmission of
sensitive information, as alerted by several non-technical online
publications~\cite{Dunn15,welive2015,Rivera2017}.

\subsection{Contribution}

In this paper, we propose an alternative to the actual Captive Portals. Our alternative
keeps their benefit (enrollment of new users with a Web
service) while exploring the existing {\WIFI} security mechanisms
defined in 802.11i and 802.11w.  This alternative explores a new
{\EAP} protocol, nicknamed {\EAPSH} (EAP for Secure Hotspots) and
relies on modified end-entities that manage the 802.1X
authentication endpoints (thus, the ones that will deal with
{\EAPSH}). To explore {\EAPSH}, users may create {\WIFI} access
profiles, just as they usually do with their home or work {\WIFI}
networks, to enforce the interaction with the correct hotspots along time.

{\EAPSH} actually combines two kinds of authentications: an
arbitrary authentication within a Captive Portal (usually
password-based) and a certificate-based authentication, using
asymmetric key pairs and public key certificates. This last method
facilitates user authentications, as it enables users to
authenticate many times during a configurable time span (the
lifetime of certificates) without providing any input after a
successful Captive Portal authentication.

For dealing with {\HTTP}-based interactions with Captive Portals,
{\EAPSH} creates a {\TLS} (Transport Layer Security~\cite{rfc8446})
tunnel over {\EAP} for transferring {\HTTP} messages between
a user browser and the Captive Portal's Web server.  This tunnel
allows the user browser to interact with a Captive Portal Web
service before the user host having an IP address (the IP
configuration will only take place after succeeding in the 802.1X
authentication protocol).

We implemented a prototype of {\EAPSH} in two open-source Linux
applications that implement the 802.1X {\EAP} endpoints:
{\WPASupplicant} and {\freeradius}. Our prototype
implementation was then successfully tested on a real setup with a
very simple, password-based Captive Portal.

This paper is structured as follows.
Section~\ref{foundations} presents the technical foundations of the 802.1X authentication.
Section~\ref{captive} presents the security issues raised by the lack of link security in hotspot networks controlled by a Captive Portal.
Section~\ref{relatedwork}
presents some related work. Section~\ref{solution} presents our {\EAP}
protocol ({\EAPSH})
and discusses some alternative exploitation scenarios.
Section~\ref{implementation} describes an implementation of {\EAPSH}
and its experimentation with a very simple Captive Portal.
Section~\ref{performance} compares the performance of {\EAPSH} with similar solutions.
Section~\ref{security_analysis} discusses the security of {\EAPSH} and
compares it with the security-flawed Captive Portal solutions.
Finally, Section~\ref{conclusions} presents the conclusions and
foresees future work.

\section{Technical Foundations}
\label{foundations}

The 802.11i standard~\cite{802.11i} defines the foundations for most
of the security mechanism used in 802.11 wireless networks. Other
standards exists, such as 802.11w~\cite{802.11w}, which is also
related with 802.11i, for complementing the security for 802.11 networks.

802.11i uses the 802.1X authentication architecture~\cite{802.1X} for performing
mutual authentication on the access of a mobile host to a wireless
network. The 802.1X architecture comprises a Supplicant (the mobile
host), an Authenticator (an Access Point, AP) and an Authentication Server (AS),
typically implemented by a RADIUS server~\cite{rfc2865}. This is commonly referred
as an enterprise architecture, because it is more suitable for
networks with a large amount of users.  In some environments,
usually referred as SOHO (Small Office, Home Office), the
architecture can be simplified and the AS is not used; this is not
our case.

\begin{figure}[h]
\center
\includegraphics[scale=.3]{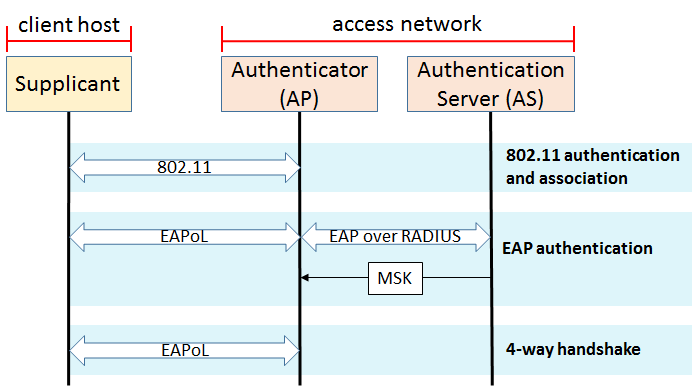}
\caption{802.1X authentication architecture: entities and phases. EAPoL stands for EAP over LAN and MSK stands for Master Session Key}
\label{802.1X.Fig}
\end{figure}

The 802.1X enterprise authentication process has three phases,
graphically displayed in Figure~\ref{802.1X.Fig}.

In the first phase, the Supplicant performs a useless network
authentication (Open System Authentication) just for fulfilling
a mandatory state-machine step inherited from WEP (Wired Equivalent
Privacy), the 802.11i predecessor, and requests its association to
the AP.

Then, we have the second phase, in which the Supplicant engages on a
mutual authentication process with the AS. This process is not
fixed; many authentication protocols can be used. However, they are
all encapsulated on EAP~\cite{rfc3748}. The
reason for using the EAP encapsulation is to allow the Authenticator
to recognize the messages exchanged in this phase, which need to be
relayed back and forth between the Supplicant and the AS. There are
many EAP authentication protocols~\cite{Dantu07};
among them, we have EAP-TLS and the one that we propose in this
paper, EAP-SH.

EAP-TLS~\cite{rfc5216} is fundamentally the establishment of a TLS~\cite{rfc8446} session with a
mutual authentication being performed with pairs of asymmetric keys
and X.509 public key certificates~\cite{rfc5280}. EAP-SH is more complex, as it can be used to enroll a user in a network provider using a Captive Portal, to fetch a
certificate for a Supplicant upon a proper authentication of its
user with a Captive Portal, and then it may work as EAP-TLS with that certificate.

The outcome of all successful EAP authentications is a Master
Session Key (MSK), which is shared between the Supplicant and the
AS. The AS uploads this key to the Authenticator, which uses it to
run the third 802.1X phase.

In the third phase, both the Supplicant and the Authenticator derive
a PMK (Pairwise Master Key) from MSK and run the so-called
four-way handshake. The purpose of this phase is to perform a mutual
authentication between Supplicant and Authenticator using PMK and to
derive a set of shorter-term traffic protection keys.

These traffic protection keys will subsequently be used to perform
the 802.11i encryption and integrity control of the data frames
exchanged between the Supplicant and the Authenticator. The 802.11i
frame integrity control assures that attackers cannot inject
false frames to any of the peers, Supplicant of Authenticator, on
behalf of the other. Moreover, neither can an attacker replay past
frames. Furthermore, the traffic protection keys will also protect
management frames, namely 802.11 authentication and association
frames, according with the 802.11w standard.

\section{Captive Portals' security issues}
\label{captive}

The list below resumes all the security problems raised by the lack of link-layer security caused by the current exploitation of hotspots with Captive Portals:
\begin{itemize}
\item Clients can abuse the network through tunnels~\cite{Prytuluk2018}
      over protocols that are always allowed, such as {\DNS}~\cite{Nussbaum09}.
\item Clients' traffic is not encrypted at the link layer, and therefore can be
      captured in clear text by radio link eavesdroppers. While some traffic is already encrypted at the application layer, this is not the case for many Internet protocols;
\item Traffic to and from clients does not have any source
      authentication or integrity control, and therefore {\MitM} and
      impersonation attacks can be performed. An example of such
      attacks are session hijacking or freeloading~\cite{Xia2004}.
      Session hijacking allows an attacker to take control of a
      session that is being held by a client host with some other
      Internet host, while freeloading allows an attacker to explore
      network resources on behalf of a client host (without its knowledge);
\item Clients and hosts, together, can be fooled by rogue hotspot
      APs, which can interfere with the clients' activities (e.g. by
      doing a {\DNS} poisoning attack~\cite{AbuNimeh2008} or browser
      history stealing~\cite{Dabrowski16}); 
\end{itemize}

The problems listed above are just an example of what can happen
when one uses a hotspot without link-layer security. Without such
security, both network clients and the network itself are open to
traffic inspection and injection. This is a very significant
security thread, which currently has no solution, not even with the
mechanisms proposed in {\WPAthree} (see Section~\ref{relatedwork}).

\section{Related Work}
\label{relatedwork}

To the best of our knowledge, there are no previous proposals
addressing the security problems created by the current paradigm for
implementing Captive Portals as we did, i.e., by using the security standards
developed for {\WIFI}, such as 802.1X, 802.11i and 802.11w.
Nevertheless, we present some other solutions, addressing the issues
raised by Captive Portals and with authentication of users accessing
hotspots.

The Universal Access Method ({\UAM}~\cite{UAM}) is a particular
implementation of the Captive Portal paradigm, providing
authentication through a Web browser. With this solution, the user
requests a Web site and the destination {\IP} address of that packet is
shifted to an {\UAM} handler. The {\UAM} handler redirects the user to the
{\UAM} portal (Captive Portal), presenting a form that the user
submits. The {\UAM} handler then contacts a {\RADIUS} server to
authenticate the user and signals the beginning of the session.

Captive Portals are ordinarily called upon clients' {\HTTP} requests, using an
{\HTTP} redirection. A new {\HTTP} status code 511~\cite{rfc6585} was
created to avoid confusions created by {\HTTP} redirections implemented
with other status codes that were being used to redirect {\HTTP}
clients (especially non-browser clients) to Captive Portals. This
solution does not solve any of the security problems created by the
Captive Portals paradigm, only helps to mitigate misinterpretations
that can arise from the redirection performed by hotspot {\APs} to
force client machines to be authenticated prior to have access to the network. 

Xia and Brustoloni~\cite{Xia2004} presented a solution for dealing
with two security vulnerabilities created by Captive Portals:
session hijacking and freeloading. Session hijacking is tackled with
automatic reauthentications performed by the client host (more
precisely, by its browser), which an attacker cannot do. Freeloading
is tackled by following sequence numbers in frames and detecting two
or more trend lines. However, the net result of both detection
mechanisms is the disconnection of the victim client host from the network. Our
solution simply does not allow session hijacking or freeloading to
take place, because it enables the communication between clients and
{\APs} to be properly protected with 802.11i~\cite{802.11i}.

Godber and Dasgupta proposed an architecture based on a security
gateway to implement an {\IPSec}~\cite{rfc6071} tunnel with the
clients' hosts upon a hotspot-based authentication~\cite{Godber02}.
This solution solves all security issues but introduces a
performance bottleneck on the security gateway, while our solution
uses the 802.11i~\cite{802.11i} protection, which distributes the workload of
wireless traffic protection by client hosts and {\APs}.

{\WIFI} hotspots are open for public access, and for that reason they
are not appropriate for exchanging authentication credentials
between the clients and the {\AP}. Some authors proposed that this
could be solved by deploying a custom developed key establishment
solution between a client host and an {\AP} using a hierarchical
identity-based cryptography set of methods~\cite{Choi11}. As with
other identity-based methods~\cite{Shamir84}, an entity public key
is derivable from a known attribute of the entity (e.g. an email
address). The corresponding private key is generated by a third
party, called the Private Key Generator ({\PKG}), based on the entity’s
identity. However, this requires the setup of a secure channel
between the entity and the {\PKG} through which it can send the private
key, which is a major disadvantage for scenarios requiring ad hoc
contacts. The {\MAC} address also serves as an identity (i.e., to
generate a public/private key pair), with the drawback of involving
all interface manufacturers (thus, raises scalability issues), and
is not retro-compatible (network interfaces need to have a private
key installed). Finally, it raises privacy issues, as the {\MAC}
address cannot be modified for preventing users' devices from being
easily traced.

Hotspot 2.0, also called HS2 or {\WIFI} Certified
Passpoint~\cite{HS2.0}, is a standard for public access {\WIFI}. It is
based on 802.11u~\cite{802.11u} and features seamless roaming
among wireless networks, as well as between {\WIFI} and cellular
networks, increased bandwidth, and services-on-demand to end-users.
This means that when a subscriber’s device is in range of at least
one {\WIFI} network, the device automatically connects to it and
network discovery, registration, provisioning and access processes
are automated. It uses an {\OSU} (Online Sign Up) {\AP} and {\OSU} server to
register future user authentication credentials (asymmetric key pair and
certified public key), which will then be used in Production {\APs} to
authenticate the client with {\EAPTLS}~\cite{rfc5216}. This solution is
somewhat similar to ours but it abolishes Captive Portals (while we
keep them) and it requires {\APs} to be modified (while we do not). 

Apple’s Captive Network Assistant ({\CNA}) is a user agent (a limited
browser) that automatically pops up and interacts with a Captive
Portal when an Apple host connects to an open {\WIFI} network of a
hotspot, instead of waiting for a browser request. The hotspot
detection is based on Apple operating systems fetching a standard
file from Apple servers, which is not served because of the
redirection to the Captive Portal~\cite{Amigopod}. A similar solution
is used by other operation systems, such as Microsoft Windows
(since their Vista version). Similar assistants exist for other
operating systems (e.g. Android), but not all assistants support all
hotspots (since there is not standard approach for implementing them).

Another alternative to notify network client hosts of the necessity to
connect to a Captive Portal is to provide its {\URI} (Uniform Resource
Identifier) within an optional field of an IPv4 {\DHCP} (Dynamic Host
Configuration Protocol) lease offer or as an extension of an IPv6
Router Advertisement message~\cite{rfc7710}. This enables clients to
trigger an interaction with the network's Captive Portal upon the
network configuration phase, instead of waiting for {\HTTP}-based
interactions initiated by the client user. This functionality,
however, has no security benefits.

The Wireless Internet Service Provider roaming
({\WISPr}~\cite{WISPr,Tan03}) is a protocol chartered by the {\WIFI}
Alliance where users are authenticated using {\UAM}. {\WISPr}
authentication is done by smart-clients (small Web browsers) which
use {\SSL}/{\XML} messaging to seamlessly login to hotspots without the
need for users to interact with the Captive Portal. {\WISPr}
authentication allows a user to roam between wireless Internet Service Providers ({\ISPs}), even if
the hotspot pertains to one {\ISP} for which the user may not have an
account. For example, if a user that has an account with {\ISP} X
attempts to roam to a hotspot of {\ISP} Y, the latter
forwards the user's credentials to X's {\AAA} (Authentication,
Authorization and Accounting) server for authentication. Once {\ISP} X
authenticates the user, it is automatically authenticated in {\ISP} Y,
as per their service agreements. {\WISPr} facilitates the sharing of
hotspots among network providers, but does not solve any of the
security issues previously presented.

WilmaGate~\cite{Brunato05} is a system for managing a single
hotspots access network with several authenticators (where
subscribers are registered). It does not solve any of the security
issues previously referred.

In~\cite{Matos12} the authors addressed the problem of
getting credentials for users, as well as the hotspot public key
certificate, using an {\NFC} (Near Field Communication~\cite{Want11})
side-channel. These credentials can then be explored to perform an
ordinary 802.1X authentication and set up a secure communication
channel using the 802.11i security mechanisms. The drawback of this
approach is that users need to have close-range contact with an {\NFC}
device to download the credentials. This approach could be used to
complement our solution for providing users with trustworthy
self-certified certificates used by hotspots.

Recently was unveiled {\WPAthree} ({\WIFI} Protected Access 3), an evolution
of {\WIFI} security standards. Among the new features included in
{\WPAthree} there is an Enhanced Open~\cite{EO18} certification, based on Opportunistic
Wireless Encryption ({\OWE}~\cite{rfc8110}). {\OWE} provides encryption
in the wireless medium but no authentication, thus it cannot tackle
active attacks by rogue {\APs}~\cite[Section~7]{rfc8110}. Furthermore,
it does not facilitates the actual authentication with
Captive Portals in any way.

\section{{\EAPSH}: Extensible Authentication Protocol for Secure Hotspots}
\label{solution}

This section describes our proposal, {\EAPSH}, which brings together
the 802.1X architecture with the Captive Portal paradigm.  {\EAPSH}
allows a client host (Supplicant) to be
redirected to a Captive Portal through {\EAP}, during which it has no
granted {\IP} access to the network. Therefore, no {\IP}-based covert
channels can be created before a successful user authentication
(e.g. {\DNS} tunneling). Furthermore, by embedding the access to a
Captive Portal within {\EAP}, and within the authentication and
authorization processes used in 802.1X, we can benefit from the key
derivation process of 802.1X, which enables the protection of the
communications between a Supplicant and an {\AP} with 802.11i protocols upon a successful
mutual authentication. Finally, this {\EAP} method was designed to
allow a transparent user authentication using previously cached
credentials, instead of relying all the time in a Captive Portal authentication.

\begin{figure*}[t]
\center
\includegraphics[scale=.3]{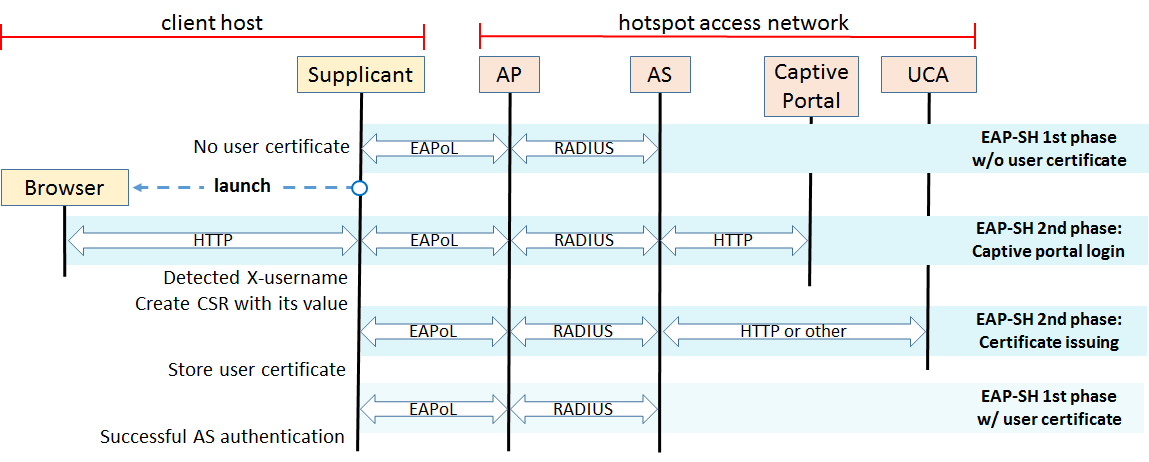}
\caption{{\EAPSH} sequence of operations for authenticating a user on  client host with
a Captive Portal within a hotspot access network. The arrows represent one or more messages and show
the protocol used in each interaction. The {\EAPoL} ({\EAP} over {\LAN})
traffic is, in this case, {\EAPSH} traffic.}
\label{eap-sh-flowchart}
\end{figure*}

\subsection{Protocol overview}

Besides the usual entities considered in the 802.1X architecture
(Supplicant, Authenticator and {\AS}), we propose
three more (see Figure~\ref{eap-sh-flowchart}): a browser, a Captive
Portal (a Web server) and a Users' Certification Authority ({\UCA}).

The browser will run in the same machine of the Supplicant and will
support the interaction with the Captive Portal (within a secure session
established between the Supplicant and the {\AS}). Note, however, that
the browser will not establish a direct {\TCP} connection with the
Captive Portal, but with the Supplicant instead.  The Captive Portal
will receive {\HTTP} messages from an {\AS}, on behalf of a Supplicant,
and will respond only to it. Finally, the {\UCA} is responsible for
issuing X.509 certificates for users upon a successful authentication on
the Captive Portal. Those certificates are accepted for user
authentication in the {\AS}, and can be cached.

We assumed that the {\AS} and the Captive Portal share a database with
user profiles and accounting information. Regarding our protocol,
the structure of such database is irrelevant.

Concerning the information shared between the ordinary 802.1X
entities and the new ones, we only need the {\AS} to know the {\UCA}
certificate (to get its public key), in order to validate 
certificates provided by users' Supplicants.

{\EAPSH} has two phases, similarly to other {\EAP}
methods~\cite{Dantu07}, such as {\PEAP}~\cite{PEAPv2} and
{\EAPTTLS}~\cite{rfc5216}. Those phases, for which high-level protocol
flows are presented in Figure~\ref{eap-sh-flowchart}, are as follows.  

In the 1st phase, there is an attempt to authenticate both
Supplicant and the {\AS} with X.509 public key certificates. The
authentication of the {\AS} cannot fail, while for the Supplicant this
can happen. If the mutual authentication succeeds, {\EAPSH} terminates
successfully.

If the Supplicant authentication could not be performed, the 2nd
phase takes place, where the user is called to perform an {\HTTP}-based
authentication with a Captive Portal. The involved {\HTTP} interaction
is relayed by the Supplicant and the {\AS}, under the
protection a {\TLS} secure session established during phase 1. Upon a
successful authentication by the Captive Portal, the Supplicant
creates a Certificate Signing Requests ({\CSR}) and sends it to the
{\UCA}, in a communication relayed by the {\AS}. Then, the {\UCA} creates a
user X.509 certificate and sends it to the Supplicant, again through
the {\AS}. Finally, the Supplicant resumes the protocol from the
beginning, in order to perform a new mutual authentication with
X.509 certificates.

For a user, the whole process involving client-side asymmetric
credentials, which are obtained in phase 2 and used in phase 1,
is completely transparent. Users are only called to interact in
authentication processes when they have to interact with the
Captive Portal, during the 2nd phase of {\EAPSH}. Thus, we keep the
Captive Portal paradigm of user registration and we allow users to
authenticate themselves with the well-known username plus password
set of credentials, but we transparently simplify the users'
authentication tasks by using the temporary asymmetric credentials.

Moreover, users are able to have multiple, simultaneous sessions
with separate authentication data, which can be revoked independently.

\subsection{{\EAP} message flows of {\EAPSH}}

An {\EAP} authentication is a request-response protocol, where requests
require the consequent reception of a response. Within 802.1X, the
initiator of a request-response dialog is always the {\AS}. The
request-response dialog terminates with either an {\EAP} success or
failure message (both without payload and unacknowledged).

\begin{figure}[t]
\center
\includegraphics[scale=.3]{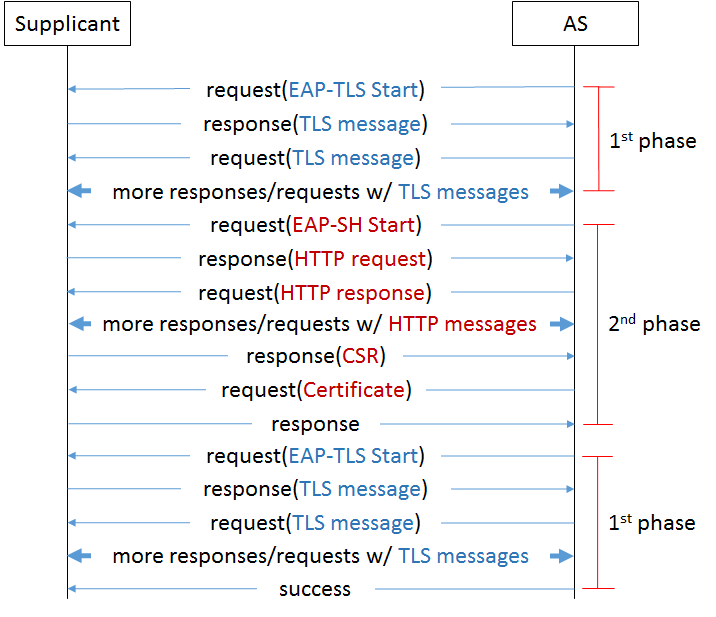}
\caption{{\EAPSH} flow of messages between the Supplicant and the {\AS}.}
\label{EAP.Fig.a}
\end{figure}

The 1st phase, or {\TLS} phase, consists on the establishment of a {\TLS}
session between the Supplicant and the {\AS} within a {\EAPTLS} protocol,
which defines a mutual authentication of the Supplicant and the {\AS}
with asymmetric key pairs and public key X.509 certificates (see
Figure~\ref{EAP.Fig.a}). However, in {\EAPSH} the Supplicant
authentication in this phase is optional, and decided by the
Supplicant, while in {\EAPTLS} it is mandatory. When the Supplicant
certificate is invalid or non-existent, this phase continues
nevertheless, until completing the setup of the {\TLS} session, and the
authentication proceeds to a 2nd phase, which runs over that secure session.

Therefore, this phase terminates with one of two {\EAP} messages:
\begin{itemize}
\item {\EAP} success: this means that the user was properly
      authenticated during phase 1, and the {\EAP} authentication, as far as
      the {\AS} is concerned, is finished.
\item {\EAPSH} Start: this means that {\EAPSH} must proceed to the 2nd phase.
\end{itemize}

The 2nd phase, or the Captive Portal phase, consists on the exchange
of {\EAPSH}-specific messages encapsulating either {\HTTP} messages, a
{\CSR} or a X.509 certificate. This phase is successful if it
terminates with the reception of a certificate by Supplicant, to be
posteriorly used in the 1st phase of {\EAPSH}, together with the
corresponding Supplicant's private key. Upon its termination, the
{\AS} initiates a new phase 1.

The success of the Captive Portal authentication is
signaled to both the {\AS} and the Supplicant with a custom {\HTTP}
header field (\texttt{X-username}) containing the user identity.
Upon receiving it, the Supplicant generates a fresh asymmetric key
pair, uses it to build a {\CSR}~\cite{rfc4211} for that identity and
sends it to the {\AS}.

The so-called ``X-'' headers have been deprecated by
RFC~6648~\cite{rfc6648}. This is not a problem, on the contrary.
Since they were deprecated, it is unlikely that our solution clashes with
the use of a similar header field, which is an advantage. On the
other hand, this header field is added by the {\AS} and removed by the
Supplicant, therefore it is never observed outside the {\EAPSH}.

When the Supplicant receives this certificate, it stores it and
re-initiates the {\EAPSH} protocol, starting again from phase 1, which
will (conceivably) authenticate the user with their asymmetric
credentials, as in {\EAPTLS}.

Upon a successful mutual authentication of both Supplicant and {\AS} at
the end of phase 1 of {\EAPSH}, the {\MSK} to be
used by both Supplicant and {\AP} to run the remainder
of the 802.1X protocol is generated just like in {\EAPTLS}.

\subsection{{\EAP} messages' structure}

{\EAP} messages have a standard header structure~\cite{rfc3748},
and a variable payload structure, depending on the specific {\EAP}
protocol being used. Therefore, we have defined our own payload for
{\EAPSH}, which is as follows.

The {\EAP} protocol does not support fragmentation of messages longer
than 1020 bytes. Since with {\EAPSH} we need to transfer possibly
longer contents between a Supplicant and an {\AS} (e.g. certificates,
Web pages, etc.), we need to handle their fragmentation on the
{\EAPSH} level.

Both requests and responses may, in some cases (e.g. when
fragmentation is used), serve as an acknowledgment of the reception
of a specific fragment.

\begin{figure}[ht]
\center
\includegraphics[scale=.3]{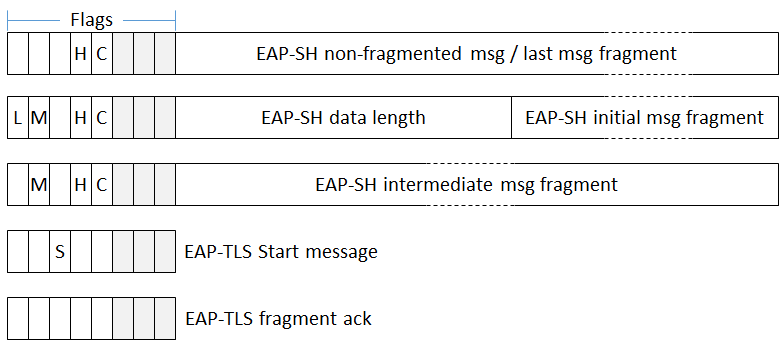}
\caption{Structure of the {\EAP} payload used for the 2nd phases of
{\EAPSH} (right). This structure was borrowed from the {\EAPTLS} message
structure. Although presented together, the \emph{H} and \emph{C}
flags are not used simultaneously; only one of them, or none, are
used at the time.}
\label{EAP.Fig.b}
\end{figure}

We use two payload structures, one for each phase. For phase 1 we
did not need to modify the {\EAPTLS} payload structure, which we
borrowed for our 2nd phase. For this, the fields of the {\EAP} payload
are the following (see Figure~\ref{EAP.Fig.b}):
\begin{itemize}
\item{\bf Flags:} 1-byte long set of bit flags. 
     The \emph{L} bit (length included) is set to indicate the presence of
     a message length field in the following bytes. This is used
     only on the first fragment of a fragmented message.
     The \emph{M} bit (more fragments) is set to indicate if more fragments
     are to come; it is active on all fragments but the last.
%
     The \emph{H} bit is set to indicate that the {\EAPSH} payload contains
     an {\HTTP} request coming from the Supplicant; for {\HTTP} responses,
     this bit is set to zero.
     The \emph{C} bit is set when the Supplicant sends its {\CSR} or
     when the {\AS} sends the Supplicant's certificate for a future {\EAPSH} phase 1.
     Finally, the \emph{S} bit ({\EAPSH} start) is set when the {\AS} initiates the 
     2nd phase.
     Fragment acknowledgments have all of their flag bits set to zero.
\item{\bf Message length:} This is is a 4-byte field with the big
     endian representation of the total length of a {\EAPSH} message.
     This length is only present on the
     first fragment of a fragmented {\EAPSH} message.
\item{\bf Payload:} This field can contain either an {\HTTP}
     message (request or response), a Supplicant’s {\CSR} or a
     Supplicant's certificate.
\end{itemize} 

\subsection{Privacy enhancement}

{\EAPTLS} is known for having privacy problems, since it exposes the
user identity in their certificate. Since the 1st phase of {\EAPSH}
is almost identical to {\EAPTLS}, we also have that problem, which
we attempted to solve.

However, privacy issues go beyond the exposure of identity labels.
In fact, usually 802.11 network devices have a fixed {\MAC} address,
which means that, even concealing more human-related attributes, an
attacker can still track devices as they move around. This was a key
motivation for {\MAC} address randomization in mobile devices, such as
smartphones~\cite{Martin17}.

The {\EAPTLS} strategy for dealing with this privacy issue consists in
repeating the protocol~\cite[\S~2.1.4]{rfc5216}: on a first step the
Supplicant provides an empty certificate list in the Certificate
Verify message, an {\AS} configured for privacy accepts the response
and continues with the setup of the secure session. Then, on a
second step, the {\AS} forces the Supplicant to run another {\TLS} session
setup, over the one previously created, but now with a proper
certificate list in the Certificate Verify message. This strategy
assumes that in the first step both peers agree on a ciphersuite
supporting confidentiality.

We considered two alternative, and more lightweight, strategies.

One possibility is to negotiate a ciphersuite supporting
confidentiality and use it to encrypt the certificate list provided
in the {\TLS} Certificate Verify message. A candidate key could be
derived from the master secret, by increasing the length of key
block that is used to derive all the required keys from the master
key~\cite[\S~2.3]{rfc5216}. Note that the master key is derived from
the pre-master key, which is provided to the {\AS} by the Supplicant
before sending the Certificate Verify message.

Another possibility is to use user pseudonyms in user certificates,
which could be renewed for each new user certificate.

In the case that no other mechanisms are in place for preventing
device tracking, such as the modification of the {\MAC} address of the
802.11 network interface on each network association, the best
strategy is the last one, the use of pseudonyms, since they do not
add any extra information to the one already gathered using only the
{\MAC} address. Otherwise, the best strategies are the first two, since
they both completely hide the details of the user certificate.

\subsection{Certificates, certification hierarchies and {\CSR}'s}

The trust anchor of the whole {\EAPSH} process is the certificate of
the {\AS} of the hotspot provider. This certificate must be trusted by
the Supplicants, directly or indirectly (by trusting some other
certificate up on its certification chain, such as a certification
root).

Note that Captive Portals do not need to have any certificate,
because browsers do not interact directly with them. Browsers
interact with the Supplicant, which interacts with the {\AS}, which
interacts with the Captive Portal. Thus, as long as the Supplicant
trusts the {\AS} certificate, the dialog between the browser and the
right Captive Portal is assured. Also note that the entire dialog between
the browser and the Captive Portal is open to inspection by the
Supplicant and the {\AS}, but encrypted while in transit between
them. A second level of security between the browser and the Web
server cannot be added as the {\AS} partially processes {\HTTP}
messages.

Regarding the user certificates, these may be generated by some
private Certification Authority ({\CA}) known only by the {\AS}, since
only the {\AS} needs to validate them. Thus, the {\UCA} does not need to
be inserted in some public certification hierarchy, it can certify
itself, since for all that matters only the {\AS} needs to trust on its
certification public key.

The actual approach for implementing pseudonyms in users'
certificates depends solely on the server side, thus they can be
defined by the hotspot network providers. From our point of view,
two strategies can be followed.

One strategy consists in having an extra column in the {\AS} users'
database with the pseudonyms. Pseudonyms may be random sequences
generated by the {\AS} upon a successful user authentication by the
Captive Portal. In this case, all the {\AS} has to do is to generate a
pseudonym once it sees an \texttt{X-username} {\HTTP} header attribute
in a Captive Portal reply, insert the pseudonym in the user's
database record and replace the value of the {\HTTP} attribute by the
pseudonym. Since the Supplicant generates a {\CSR} with whatever comes
in the \texttt{X-username} attribute, it will use the pseudonym
transparently.

Another strategy consists in doing the same, but using a
pseudonym computed by some encryption performed internally by the {\AS}.
In this case, the users' database does not need to have any extra
column, since the {\AS} can get the correct user identity by decrypting
the user pseudonym. Different pseudonyms for the same user can be
easily achieved by using some random data in the encryption process
(e.g. encrypting with a secret key and a random Initialization Vector ({\IV}) or encrypting with the {\AS} public key with {\PKCS}~\#1 or
{\OAEP} (Optimal Asymmetric Encryption Padding) paddings~\cite{rfc8017}).

For coping with any identity value provided by an {\AS}, assuming that
it is provided as a {\ASCII} string, a Supplicant can simply use it as
the Common Name in the Subject field of a {\CSR}.

\subsection{Dialog with the Captive Portal}

The dialog between a user and a Captive Portal uses a browser, as
usually. However, the browser is automatically launched by the
Supplicant application only when necessary, i.e., when the
authentication cannot be performed using asymmetric credentials. A
simplified, or a more secure browser, can be used if necessary,
and this is decided by the client or the Supplicant provider.

The browser connection is directed to a local {\URI}, i.e., to the
localhost Internet address and to a {\TCP} port allocated by the
Supplicant application. This port receives the {\HTTP}
traffic produced by the browser, which is then forwarded to the
Captive Portal through the {\AS}, which also establishes a {\TCP}
connection with the Captive Portal for that purpose. The reverse
path of that {\HTTP} traffic is treated similarly: the {\AS} receives the
responses from the Captive Portal, forwards them to the Supplicant,
and this forwards them again to the browser. Therefore, the objects
in the Web page (e.g. {\CSS}, JavaScript, images) should be added by taking in 
consideration that resources are available to the browser at the localhost,
and not at the Web server {\IP} address.


The browser does not need to authenticate the Captive Portal, as in
other solutions using Captive Portals. In fact, the network provider
authentication is tackled in the establishment of the {\TLS} tunnel,
thereafter all the dialog between a browser and the Captive Portal
is guaranteed to be performed with the correct entity because it
goes through the {\TLS} tunnel.

\subsubsection{Password management issues}

Our transparent proxying between the user browser and the Captive
Portal raises some issues related with the storage of login
credentials by browsers' password managers. Today, and unless
configured otherwise, browsers are extremely effective in helping
users to memorize login credentials, usually spotted when {\HTML}
forms use an input field with the \verb|password| type.

Browsers' password managers memorize login credentials associated to
the {\URL} portion that provides the service endpoint, i.e., its host
name (or {\IP} address) and {\TCP} port. However, in our case we cannot
guaranty that the Supplicant always uses the same {\TCP} port to handle
browsers' dialogues, unless a specific port could be allocated by
{\IANA}\footnote{Internet Assigned Numbers Authority,
\url{http://www.iana.org}}. Furthermore, even if using the exact
same port, the constancy of the {\IP} address (localhost) would merge
credentials for different {\EAPSH} hotspots under the same memory
record, in which case it may create problems if using the same
username across different hotspot providers.  

Concluding, it should not be easy to use browser's password managers
to memorize users' credentials for accessing {\EAPSH} Captive
Portals.  The obvious option is, then, not to use such managers, and
always input the login credentials in Captive Portals' interfaces.
The non-memorization of credentials can be forced by Captive
Portals, but this requires replacing simple and straightforward
login {\HTML} forms by JavaScript code capable of doing the same (as
we did in our implementation, see Section~\ref{CP}). If such
non-memorization is not forced, users must voluntarily reject the
memorization of credentials each time they log in a Captive Portal,
unless the browser provides some permanent disabling of that (which
is only effective if the same local {\URL} is presented to the browser).

\subsection{Validation of the {\AS} certificate}

Usually certificates provided by servers, such as the {\AS}, are
validated by the Supplicant using certification hierarchies and
certificate revocations. However, for checking certificate
revocations one cannot resort to {\OCSP} (Online Certificates
Status Protocol~\cite{rfc6960}) servers, because the Supplicant at this
point cannot use {\IP} communications, and it is not practical to use
Certificates Revocation Lists, because that would imply a
pre-caching of all the required lists before performing a first
connection with the {\AS}.  

The solution for this problem is to use the so-called {\OCSP} stapling
mechanism~\cite{rfc6066}. A {\TLS} client may require, on its initial
Hello message, the provisioning by the server of a signed {\OCSP}
message with the current revocation status of the server's
certification chain. When the client appends to its Hello message a
Certificate Status Request, the server sends, along with its
certificate, a status field with an up-to-date {\OCSP} response. This
response allows the Supplicant to validate the server's certificate
without having to initiate connections with {\OCSP} servers.

\section{Prototype Implementation}
\label{implementation}

For evaluating the suitability of {\EAPSH} we implemented it using
existing open-source applications for Linux systems. For the Supplicant we
modified the {\WPASupplicant} application; for the {\AS} we modified a
{\freeradius} server.

Regarding the {\UCA} and the related certification services, we used a
private {\PKI} managed with calls to the {\openssl} application. This
application was also used by the Supplicant to create a {\CSR} and by the {\AS}
(on behalf of a {\UCA}) to issue a user certificate from a {\CSR}.

Finally, the Captive Portal was implemented with a
CherryPy\footnote{\url{http://cherrypy.org}} Web server.

This prototype was successfully used to perform consecutive logins
with a Linux client host.

%
%
%
%

\subsection{Supplicant}

On Linux, the 802.1X Supplicant is played by a daemon that controls
the 802.11 wireless networking, called {\WPASupplicant}, which
supports a wide variety of {\EAP}
methods~\footnote{\url{https://linux.die.net/man/8/wpa_supplicant}}.
Each {\EAP} method is implemented as a separated module and all {\EAP}
methods use the same interface between the supplicant state machine
and method-specific functions. Method registration is done through
this interface. This allows new {\EAP} methods to be added without
modifying the core {\EAP} state machine implementation.

Regarding this core implementation, all we had to do was to add a
new {\EAP} type number to the set of {\EAP} type definitions. As of today,
numbers $56$ to $191$ are
unassigned~\footnote{\url{http://www.iana.org/assignments/eap-numbers/eap-numbers.xhtml}},
therefore we assigned to {\EAPSH} the type number $56$.  

As shown before, our {\EAPSH} incorporates an {\EAPTLS} phase, similarly
to {\PEAP} or {\EAPTTLS}~\cite{rfc5281}. For that reason, we adapted the
{\WPASupplicant}’s {\PEAP} module, which also uses {\TLS} on a 1st stage,
for implementing {\EAPSH}.

The {\TLS} component of {\WPASupplicant} is generic enough to be used
alone or combined with other {\EAP} methods. Namely, it supports client
authentication (used by {\EAPTLS}) or is able to proceed without it.
This was a perfect match for our {\EAPSH}.  

\subsubsection{Configuration}

The {\WPASupplicant} configuration is given by a text file (usually
\verb|wpa_supplicant.conf|) that lists all accepted networks and
their security policies. It can include one or more network
definition blocks, which include the entire necessary configuration for
those networks. 

For {\EAPSH}, the network block comprises a number of fields 
that contain critical information used when authenticating to a
network, which are the following:
 
\begin{itemize}
\item \verb|ssid|: Sequence of characters that uniquely identify a {\WLAN};
\item \verb|key_mgmt|: Accepted authenticated key management
      protocol (\verb|WPA-EAP| for {\EAPSH});
\item \verb|eap|: {\EAP} method used when authenticating to this
      network (\verb|{\EAPSH}|, in our case);
\item \verb|ca_path|: path of the directory containing
      {\CA} certificates files to validate the network's {\AS}
      certificate;
\item \verb|client_cert| and \verb|private_key|: Paths to
      the client's certificate file and corresponding private key
      file, respectively.
      The contents of these files, if valid, are used in the 1st
      phase of {\EAPSH}; every complete run of its 2nd phase 
      updates their contents.
\item \verb|private_key_password|: The password used to protect the
      user private key used in the 1st phase of {\EAPSH}.
\item \verb|browser_command|: The command to launch a browser to
      interact with the Captive Portal (through a local socket owned
      by {\WPASupplicant}). We used the \verb|firefox -new-window %s|
      command to launch the browser with a given {\URL}, which
      would replace the parameter string \verb|%s|.
\end{itemize}

Unlike for similar configurations, the user does not need to specify
the \verb|ocsp| field for requiring {\OCSP} stapling, as {\EAPSH} does
that by default.

\subsubsection{{\EAPSH} module on {\WPASupplicant}}

This module starts by instantiating a {\TLS} tunnel, using or not the
user authentication with asymmetric credentials. At the end of
this step the {\AS} sends an {\EAP} message stating whether or not the
user was authenticated during the setup of the {\TLS} tunnel. This
message has an empty payload; the information is solely coded in its
\emph{S} flag bit. If set, the {\WPASupplicant} proceeds to the 2nd
phase; otherwise, it terminates successfully.

For the 2nd phase, the {\WPASupplicant} starts by allocating a
socket and binding it to a random, local {\TCP} endpoint (i.e., with an
127.X.X.X {\IP} address and a port between 1025 and 65535). Then,
it launches the user browser (as configured by the user) for
accessing the {\URL} constituted by the random {\TCP} endpoint.

Upon this initial setup, the {\WPASupplicant} enters the following
processing cycle:
\begin{enumerate}
\item Read a full {\HTTP} request from its {\TCP} endpoint;
\item Transform the {\HTTP} request into a {\TLS} encrypted message and
      store it on one or more {\EAPSH} messages with the \emph{H} bit set;
\item Send the {\EAPSH} message to the {\AS};
\item Receive {\EAPSH} messages with the \emph{H} bit set from the
      tunnel, combine them until having a complete {\HTTP} response, and send
      it to the {\TCP} endpoint.
\item Search for an \texttt{X-username} field in the header of the {\HTTP}
      response. If present, terminate this cycle and shutdown
      the {\TCP} endpoint.
\end{enumerate}

Once knowing the user pseudonym that should be used in their
certificate, the {\WPASupplicant} proceeds to create a new {\RSA} key
pair and a corresponding {\CSR}. Then, the {\WPASupplicant}
encapsulates the {\CSR} on a {\TLS} message, stores it on one or more
{\EAPSH} messages with the \emph{C} bit set and sends them to the {\AS}.

Finally, the {\WPASupplicant} waits for a final {\EAPSH} message from
the {\AS}, which must contain the user's public key certificate. It
stores the certificate in a file with the path given by the
configuration file and starts the whole process again, this time in
the hope of terminating at the end of phase 1.

For this second round of {\EAPSH}, the {\TLS} tunnel must be recreated
from scratch, i.e., the previous {\TLS} session cannot be reused. This
is required to force the client authentication, which cannot take
place when a {\TLS} session is resumed.

\subsubsection{{\CSR} generation}

The {\CSR} is generated with the {\openssl} tool, using a configuration
file created by {\WPASupplicant}. The contents of that file are
dynamic (i.e. the file does not need to pre-exist) and presented in
Figure~\ref{CSR.Fig}.
The Subject's Common Name is filled with the
identity provided by the {\AS}, interpreted as an {\ASCII} string. The key
pair used in the {\CSR} is always a new one, with 1024 bits.

\begin{figure}[t]
\begin{lstlisting}[frame=single,basicstyle=\scriptsize\ttfamily]
[ req ]
prompt=no
default_keyfile = "provided in the user's conf file"
output_password = "provided in the user's conf file"
default_bits = 1024
default_md = sha256
distinguished_name = DN

[ DN ]
CN = "pseudonym provided by the AS"
\end{lstlisting}
\caption{Configuration file provided to {\openssl} to generate a user's {\CSR}. Part of its contents are filled with data collected from the {\WPASupplicant} configuration file}
\label{CSR.Fig}
\end{figure}

\subsection{{\RADIUS} server}

The authentication server was implemented by a modified {\RADIUS}
server. {\RADIUS}~\cite{rfc2865} is a protocol that provides
centralized Authentication, Authorization and Accounting management
for users that connect and intend to use a network service. We
decided to modify {\FreeRADIUS}\footnote{\url{http://freeradius.org/}},
a popular open source {\RADIUS} server, to implement the {\AS} component
of {\EAPSH}.

 
Similarly to {\WPASupplicant}, adding {\EAPSH} to {\freeradius}
implied creating a set of C files implementing the {\EAP} interface
methods (such as \texttt{instantiate} and \texttt{process}), and the
protocol's logic. As for {\WPASupplicant}, we adapted the
{\freeradius}'s {\PEAP} modules to implement {\EAPSH}.

\subsubsection{{\HTTP} handling and pseudonym generation}

{\HTTP} requests, once defragmented from {\EAP} messages, are sent to the
Captive Portal using a new socket connection. Then, the {\EAPSH}
module waits for the {\HTTP} response and, once complete, searches its
header for the \texttt{X-username} field.

\newcommand{\conc}{\,\|\,}
If present, it encrypts its value with {\AES}-128 in {\CBC} mode with
{\PKCS}~\#7 padding and a random {\IV}. Then, the
{\IV} is concatenated with the resulting cryptogram, is authenticated
with an {\HMAC}~\cite{rfc2104} computed with {\SHA}-1~\cite{rfc3174} and the result is encoded in Base64, yielding the
user's pseudonym:
$$
c = \text{{\AES}-128-{\CBC}}\left( K, \text{\IV}, \text{identity} {\conc} \text{{\PKCS} \#7 padding} \right) \\
$$
$$
\text{pseudonym} = \text{Base64}\left( \text{\IV} {\conc} c {\conc} \text{\HMAC}\left( \text{\SHA-1}, K, \text{{\IV}} {\conc} c \right) \right)
$$
where $K$ is the encryption and authorization key and ${\conc}$
denotes concatenation. The random {\IV} assures that different
pseudonyms will be generated for the same identity on different
occasions.

The same key is used to encrypt the username and check for the
integrity of a pseudonym. In our implementation, this key is volatile
and generated each time the {\AS} is launched. The consequence of its
loss upon a server restart is that users need to fetch a new
certificate, thus need to be authenticated by the Captive Portal.

The pseudonym is then used to replace the former value of the
\texttt{X-username} field. Finally, the pseudonym is stored is a
short term, volatile cache, for validating subsequent certification
requests.  

The pseudonym on a user certificate can be decrypted upon a
successful 1st phase of {\EAPSH}, yielding the real username. This
username can them be used for other operations, such as logging or
authorization. We did not implement them in our prototype.

\subsubsection{{\CSR} handling}

When a {\CSR} is received, its Subject's Common Name is fetched from it
and checked against the pseudonyms cache. If present and still fresh
(we used a 1-minute time window), the {\CSR} is used to generate a new
user's certificate.

To simplify the implementation for the proof of concept, the {\EAPSH}
module also acts as a {\UCA}. This actual certificate generation is
performed by a shell script launched by the {\EAPSH} module,
which uses the {\openssl} command. All this command's parameters,
such as the validity period of the users' certificates, are defined
by this script. By using a script file, we are able to tune the
parameters used in the certification process without having to
change the {\freeradius} source code nor to add all the
certification options to {\freeradius} configuration files.

\subsubsection{Certificates and certification chains}

For this prototype, we implemented a very simple {\PKI}, formed by a
root {\CA} and an intermediate {\CA}. The intermediate {\CA} was used to
issue the certificate used by the {\EAPSH} module of {\RADIUS} server and
to issue client's certificates, upon reception of their {\CSR}. To
manage all the aspects of this {\PKI} we also used {\openssl}.

\subsubsection{{\EAPSH} module configuration}

In order to integrate our protocol, the {\freeradius} 
{\EAP} modules configuration file was changed accordingly.
This file is parsed by the {\EAPSH} module upon its initialization and its contents are the following:

\begin{itemize}
\item \verb|certificate_file|: file with the {\AS} certificate
      and certification chain;
\item \verb|private_key_file|: file with the {\AS} private key;
\item \verb|private_key_password|: password to decrypt the private
      key file;
\item \verb|ca_file|: server {\CA} file, for validating 
      Supplicants' certificates;
\item \verb|user_cerficate_issuing_script|: shell script used to
      issue a user certificate;
\item \verb|captive_portal_endpoint|: the {\IP}
      address and {\TCP} port of the Captive Portal {\HTTP} server.
\end{itemize}

\subsection{Captive Portal}
\label{CP}

This server was implemented with a CherryPy Web server,
implementing a very simple authentication page based on a username
and a password.  Its user database was implemented with
{\SQLite}\footnote{\url{https://www.sqlite.org}}.

The login page uses an {\HTML} form that was designed with
the goal of preventing browsers from memorizing users' credentials.
Empirically we found out that browsers are extremely effective in
such memorization, which is triggered by the presence of an input
field of type \texttt{password}.

Therefore, we replaced the \texttt{password} input type by
\texttt{text} and used some extra JavaScript event handlers to hide the
password, character by character, immediately after its input
(see Figure~\ref{Login.Fig}).
The original password characters are saved
in a variable, which is copied (without any hashing) into a hidden form field when the
latter is submitted. With this approach, we managed to disable the
storage of user's passwords by the browsers' password manager while
hiding the password typed from shoulder surfers.

The fact of not using a form input field of type \texttt{password}
has another interesting side effect. Usually browsers warn users
when they fill credential forms belonging to HTTP resources, because
those forms will be exposed in the network on their submission. The way
browsers detect the input of sensitive data, such as credentials, in
forms is by the presence of fields of type \texttt{password}. Thus,
because we did not use this field type, for the reasons presented
above, we also avoid useless browser warnings (because, in our case,
the browser session connects to a local port that forwards the HTTP
traffic to a Captive Portal over a TLS tunnel; therefore, there is no risk).

If, in the future, browsers start imposing HTTPS-only access, there
are two possible solutions: one is to disable (or not to implement
at all) this feature for traffic that is internal to the machine; a
second one is to provide an HTTPS endpoint on the Supplicant, using
a certificate belonging to a certification chain used only internally.
Naturally, the root of that certification chain would have to belong
to the list of trusted roots used by local browsers.

\begin{figure}[ht]
\begin{lstlisting}[frame=single,basicstyle=\scriptsize,language=html]
<script>
var psw = "";
var star = "";

function hide() {
  var v = document.getElementById("psw").value;
  if (v.length < star.length) {
    psw = psw.substring(0, v.length-1);
    star = star.substring(0, v.length-1);
  } else {
    star += "*";
    v = document.getElementById("psw").value;
    psw += v.slice(-1);
  }
  document.getElementById("psw").value = star;
}

function unhide() {
  document.getElementById("hpsw").value = psw;
}
</script>

<form onsubmit=unhide() action="/login"
      method=post>
  <input type="hidden" name="hpsw" id="hpsw"
         autocomplete="off">
  Username<br>
   <input type="text" name="uname" required
          autocomplete="off"><br>
  Password<br>
   <input type="text" name="psw" id="psw" required
          autocomplete="off" oninput=hide()><br>
  <button type="submit">Login</button>
</form>
\end{lstlisting}
\caption{{\HTML} login form, with JavaScript event handlers to save and hide
the password as it is typed and to provide the saved password
(without any post-processing, such as hashing) when the form is
submitted (in its hidden input field ``hpsw'').}
\label{Login.Fig}
\end{figure}

\section{Performance evaluation}
\label{performance}

The performance of our solution depends on many technical aspects,
many of them highly variable, such as the performance of the
machines involved, the performance of the communications and the
cost of the (asymmetric) cryptographic operations involved. In addition,
it depends on the specific interaction with a
Captive Portal and on pauses due to the required human-machine
interaction. Therefore, we decided not to evaluate the performance
of our solution in absolute terms, but rather to compare it
theoretically with similar solutions.

The first time a user enters a hotspot network they need to run the
complete protocol displayed on Figure~\ref{EAP.Fig.a}. The first and
the second phases, together, are similar to the actual
authentication with a Captive Portal. In fact, Captive Portal
authentications typically run over an {\HTTPS} session (to protect
the confidentiality of the credentials entered by users). Thus, our
(initial) 1st phase is similar to the set-up of a {\TLS} session
within an {\HTTPS} interaction, and our 2nd phase is similar to the
{\HTTP} interaction with a Captive Portal that runs within {\HTTPS}.
In both cases, the only differences are the {\EAP} and {\RADIUS}
encapsulation displayed in Figure~\ref{eap-sh-flowchart} (which add
a small overhead on the exchanged data), and the generation and
exchange of the {\CSR} and user certificate at the end of the second
phase.

For users that already have the asymmetric credentials fetched with
the interactions previously referred, the authentications are equal
to {\EAPTLS} (final first phase in Figure~\ref{EAP.Fig.a}). Since this
is intended to be the most frequent case, because the asymmetric
credentials can be valid for some time (e.g.  a day), {\EAPSH} will
most of the time be as efficient as {\EAPTLS}.

Concluding, the only relevant overhead introduced by {\EAPSH}, when
comparing with the actual authentication protocols used in {\WIFI}
networks (including with Captive Portals), is the sporadic
generation and exchange of a {\CSR} and the corresponding
certificate. Just to illustrate how small can this overhead be, both
the generation of a {\CSR} (with a new $1024$-bit RSA key) and the
generation of a certificate (signed by a $2048$-bit RSA key) take
around $100$ milliseconds each in a Windows 10 host with an Intel 
core i7-8550U processor.

\section{Security and Usability Analysis}
\label{security_analysis}

With {\EAPSH}, the hotspot provider is authenticated with a
certificate that needs to be pre-configured by the Supplicant. This
means that, once such configuration has been done, the user does not
need to pay attention to such action again (except if the
certificate changes). This is not the case with the current Captive
Portal authentication, where a user needs to pay attention to
the effective security of every {\HTTPS} session established
with each Captive Portal.

The import of the hotspot certificate for the client hosts is,
therefore, a critical task. Today this can be done in many different
ways. For instance, the certificate may be downloaded from a secure
Web page, using an existing network access, prior to use a hotspot
(e.g. using a cellular interface). Alternatively, the certificate can be decoded
from a {\QR} code image displayed in a secure place (e.g. painted in
the wall of a cybercafe), or downloaded
from a distribution device using a wireless communication technology
immune to {\MitM} attacks (e.g. {\NFC}, as in~\cite{Matos12}). In short,
there are numerous secure ways to convey the hotspot certificate
to its clients, but the discussion of their relative benefits is out
of the scope of this paper. For what concerns us, this is not a
complex problem to solve.

Although not previously referred, the {\EAPSH} paradigm can help users
to identify when they are accessing a Captive Portal with it. In
fact, the {\EAPSH} Supplicant software has access to the {\HTML} contents
provided by a Captive Portal, and can easily manipulate them in
order to add to it some personal, visual content available only
locally and customized by users. This way, a rogue Captive Portal
Web page could not impersonate thz ones effective observed by each
user for all {\EAPSH} based Captive Portals.

The {\EAPSH} approach extends the 802.1X paradigm. This means that it
enables a secure interaction between the Supplicant and the {\AP} upon
a successful authentication between the Supplicant and the {\AS}. This
fact rules out some of the already referred attacks, such as session
hijacking or freeloading~\cite{Xia2004} or {\DNS}
poisoning~\cite{AbuNimeh2008}. This is not possible with the current
Captive Portal approach, where the wireless communication is never
protected at the link layer.

The {\EAPSH} protects the hotspot provider from attacks deployed by
non-authenticated users. With the current Captive Portal paradigm,
client hosts need to configure their network layer (get an {\IP}
address and a netmask), for which they need to run a {\DHCP}
negotiation. Many times, they are even allowed to perform {\DNS}
queries, which enables them to contact networks beyond the one of
the hotspot provider (using, for instance, {\DNS}
tunnels~\cite{Nussbaum09}). All these possibilities
represent a security risk, since such services can be abused and
there is no way to identify the responsible person. On the contrary,
with {\EAPSH} the client host does not have {\IP} access to the network
until successfully finishing its 802.1X authentication.  

The {\CSR} created by the Supplicant uses a user name (or pseudonym)
that was given to the Supplicant and that is known by the {\AS}. 
Therefore, a malicious Supplicant, once passed the Captive Portal
authentication, cannot impersonate other users by using their
pseudonym. This check is done when the {\AS} receives the {\CSR}, before
forwarding it to the {\UCA}.

\section{Conclusion and Future Work}
\label{conclusions}

In this paper, we proposed a new {\EAP} protocol, {\EAPSH}, which is able
to merge two authentication paradigms: the one using Captive Portals
and the one of 802.1X. The first is useful for enrolling new clients
by a hotspot provider, and the second is able to provide security
to the {\WIFI} link between a client host and the hotspot {\AP}.
The current Captive Portal
paradigm is a hack that requires a special support by the {\APs} and
does not provide any security on the wireless link. On the contrary,
{\EAPSH}, combined with 802.1X, requires the modification of the
Supplicant and {\AS} systems, but works with all existing {\APs} that
support the 802.1X enterprise mode, which is common nowadays. 

{\EAPSH} not only provides support for a Captive Portal authentication
but it also enables client hosts to authenticate consecutively in
the hotspot network without requiring users to provide
authentication credentials (usually a username and password pair).
This is achieved with the just in time issuing of client
certificates, which may be cached by client hosts and used during
the certificates' lifetime. The only drawback of our approach is the
fact that users are not able to memorize their Captive Portal's
credentials using the browsers' password management facilities,
since there is no way to associate the right credentials to each
hotspot provider. Nevertheless, the possibility of using the
certificates to authenticate eases this problem, since the number of
required Captive Portal authentications is reduced.

In terms of added security, {\EAPSH} clearly outperforms the current
Captive Portal paradigm. Client hosts cannot use {\IP} networking prior
to complete their authentication, thus they are not able to use the
network before completing their authentication. Hotspot providers
can be authenticated with certificates associated with {\EAPSH}
profiles, thus users are more protected from attacks by rogue
hotspot providers. Finally, the wireless link between the client
host and the hotspot {\AP} is secured (encrypted
and authenticated with 802.11i protocols) upon a successful authentication.

A prototype of {\EAPSH} was implemented in two Linux tools that
implement the Supplicant and the {\AS} of an 802.1X architecture:
{\WPASupplicant} and {\freeradius}. This prototype was
successfully tested with a simple password-based Captive Portal,
which was designed for avoiding clients' browsers to use their
password manager to store users' credentials.

As future work, we can foresee several lines of action. A first one
is to pursue the standardization of {\EAPSH} in the form of an {\IETF}
{\RFC}. A second one is a closer integration of the Supplicant software
with a browser for enabling the memorization of users' passwords by
the browsers' password manager, a facility that is used by many
users. Finally, a third one, also related with that closer
integration, the presentation of a custom user interface provided by
the Supplicant (e.g. a background image) to assure users that they
are interacting with a given hotspot provider when they are called
to provide their credentials to a Captive Portal.

 


\bibliography{main,rfc}

\begin{thebibliography}{10}

\bibitem{rfc3748}
B.~Aboba, L.~Blunk, J.~Vollbrecht, J.~Carlson, and H.~Levkowetz.
\newblock {Extensible Authentication Protocol (EAP)}.
\newblock RFC 3748 (Proposed Standard), June 2004.

\bibitem{AbuNimeh2008}
S.~Abu-Nimeh and S.~Nair.
\newblock {Bypassing Security Toolbars and Phishing Filters via DNS Poisoning}.
\newblock In {\em IEEE Global Telecommunications Conf. (IEEE GLOBECOM 2008)},
  New Orleans, LA, USA, November 2008.

\bibitem{HS2.0}
Wi-Fi Alliance.
\newblock {Hotspot 2.0 (Release 2) Technical Specification}, 2016.

\bibitem{Brunato05}
M.~Brunato and D.~Severina.
\newblock {WilmaGate: A New Open Access Gateway for Hotspot Management}.
\newblock In {\em Proc. of the 3rd ACM Int. Workshop on Wireless Mobile
  Applications and Services on WLAN Hotspots (WMASH '05)}, pages 56--64,
  Cologne, Germany, 2005.

\bibitem{Choi11}
J.~Choi, SY. Chang, D.~Ko, and YC. Hu.
\newblock {Secure MAC-layer protocol for captive portals in wireless hotspots}.
\newblock In {\em 2011 IEEE Int. Conf. on Communications (ICC)}, pages 1--5.
  IEEE, 2011.

\bibitem{Cisco2019}
Cisco.
\newblock {Cisco Visual Networking Index: Global Mobile Data Traffic Forecast
  Update, 2017-2022 White Paper}, February 2019.
\newblock
  \url{http://www.cisco.com/c/en/us/solutions/collateral/service-provider/visual-networking-index-vni/mobile-white-paper-c11-520862.pdf}.

\bibitem{rfc5280}
D.~Cooper, S.~Santesson, S.~Farrell, S.~Boeyen, R.~Housley, and W.~Polk.
\newblock {Internet X.509 Public Key Infrastructure Certificate and Certificate
  Revocation List (CRL) Profile}.
\newblock RFC 5280 (Proposed Standard), May 2008.
\newblock Updated by RFC 6818.

\bibitem{Dabrowski16}
Adrian Dabrowski, Georg Merzdovnik, Nikolaus Kommenda, and Edgar Weippl.
\newblock {Browser history stealing with captive Wi-Fi portals}.
\newblock In {\em IEEE Security and Privacy Workshops (SPW 2016)}, pages
  234--240, San Jose, CA, USA, May 2016.

\bibitem{Dantu07}
R.~Dantu, G.~Clothier, and A.~Atri.
\newblock {EAP methods for wireless networks}.
\newblock {\em Computer Standards \& Interfaces}, 29(3):289--301, 2007.

\bibitem{Dunn15}
John~E. Duhn.
\newblock {Are public Wi-Fi hotspots a security risk? Security risks of using
  public Wi-Fi explained}.
\newblock Computerworld UK, August 2015.
\newblock
  \url{https://www.computerworlduk.com/security/are-public-wi-fi-hotspots-really-major-security-risk-3623447}.

\bibitem{rfc6066}
D.~{Eastlake 3rd}.
\newblock {Transport Layer Security (TLS) Extensions: Extension Definitions}.
\newblock RFC 6066 (Proposed Standard), January 2011.

\bibitem{rfc3174}
D.~{Eastlake 3rd} and P.~Jones.
\newblock {US Secure Hash Algorithm 1 (SHA1)}.
\newblock RFC 3174 (Informational), September 2001.
\newblock Updated by RFCs 4634, 6234.

\bibitem{rfc6071}
S.~Frankel and S.~Krishnan.
\newblock {IP Security (IPsec) and Internet Key Exchange (IKE) Document
  Roadmap}.
\newblock RFC 6071 (Informational), February 2011.

\bibitem{rfc5281}
P.~Funk and S.~Blake-Wilson.
\newblock {Extensible Authentication Protocol Tunneled Transport Layer Security
  Authenticated Protocol Version 0 (EAP-TTLSv0)}.
\newblock RFC 5281 (Informational), August 2008.

\bibitem{Godber02}
A.~Godber and P.~Dasgupta.
\newblock {Secure Wireless Gateway}.
\newblock In {\em Proc. 1st ACM Workshop on Wireless Security (Wise '02)},
  pages 41--46, Atlanta, GA, USA, 2002.

\bibitem{rfc8110}
D.~Harkins and W.~Kumari.
\newblock {Opportunistic Wireless Encryption}.
\newblock RFC 8110 (Informational), March 2017.

\bibitem{UAM}
i~Sprint.
\newblock {AccessMatrix UAM: common security platform for enterprise
  applications}, 2000.
\newblock \url{http://www.i-sprint.com/brochure/uam_en.pdf}.

\bibitem{rfc2104}
H.~Krawczyk, M.~Bellare, and R.~Canetti.
\newblock {HMAC: Keyed-Hashing for Message Authentication}.
\newblock RFC 2104 (Informational), February 1997.
\newblock Updated by RFC 6151.

\bibitem{rfc7710}
W.~Kumari, O.~Gudmundsson, P.~Ebersman, and S.~Sheng.
\newblock {Captive-Portal Identification Using DHCP or Router Advertisements
  (RAs)}.
\newblock RFC 7710 (Proposed Standard), December 2015.

\bibitem{Martin17}
J.~Martin, T.~Mayberry, C.~Donahue, L.~Foppe, L.~Brown, C.~Riggins, E.~C Rye,
  and D.~Brown.
\newblock {A Study of MAC Address Randomization in Mobile Devices and When it
  Fails}.
\newblock {\em arXiv preprint arXiv:1703.02874}, 2017.

\bibitem{Matos12}
A.~Matos, D.~Romão, and P.~Trezentos.
\newblock {Secure hotspot authentication through a Near Field Communication
  side-channel}.
\newblock In {\em IEEE 8th Int. Conf. on Wireless and Mobile Computing,
  Networking and Communications (WiMob)}, pages 807--814, Oct 2012.

\bibitem{rfc8017}
K.~Moriarty, B.~Kaliski, J.~Jonsson, and A.~Rusch.
\newblock {PKCS \#1: RSA Cryptography Specifications Version 2.2}.
\newblock RFC 8017 (Informational), November 2016.

\bibitem{Amigopod}
Aruba Networks.
\newblock {Apple Captive Network Assistant Bypass with Amigopod, Version 1.0},
  2012.
\newblock
  \url{https://www.arubanetworks.com/vrd/ACNAAppNote/wwhelp/wwhimpl/js/html/wwhelp.htm}.

\bibitem{EO18}
Aruba Networks.
\newblock {WPA3 and Enhanced Open: Next generation Wi-Fi security}.
\newblock White Paper, 2018.
\newblock
  \url{https://www.arubanetworks.com/assets/wp/WP_WPA3-Enhanced-Open.pdf}.

\bibitem{Nomadix2013}
Nomadix.
\newblock {Captive portal and the new security paradigm: Options for handling
  redirection problems caused by certificate mismatches}, 2013.
\newblock
  \url{https://www.anixter.com/content/dam/Suppliers/nomadix/SSL_Redirection.pdf}.

\bibitem{rfc6585}
M.~Nottingham and R.~Fielding.
\newblock {Additional HTTP Status Codes}.
\newblock RFC 6585 (Proposed Standard), April 2012.

\bibitem{Nussbaum09}
L.~Nussbaum, P.~Neyron, and O.~Richard.
\newblock {On robust covert channels inside DNS}.
\newblock {\em Emerging Challenges for Security, Privacy and Trust}, pages
  51--62, 2009.

\bibitem{802.11i}
LAN/MAN Standards~Committee of~the IEEE Computer~Society.
\newblock {Wireless LAN Medium Access Control (MAC) and Physical Layer (PHY)
  Specifications, Amendment 6: Medium Access Control (MAC) Security
  Enhancements}.
\newblock IEEE Std 802.11i, July 2004.

\bibitem{802.11w}
LAN/MAN Standards~Committee of~the IEEE Computer~Society.
\newblock {Wireless LAN Medium Access Control (MAC) and Physical Layer (PHY)
  Specifications, Amendment 4: Protected management Frames}.
\newblock IEEE Std 802.11w, July 2009.

\bibitem{802.1X}
LAN/MAN Standards~Committee of~the IEEE Computer~Society.
\newblock {IEEE Standard for Local and Metropolitan Area Networks: Port-Based
  Network Access Control}.
\newblock IEEE Std 802.1X-2010, February 2010.

\bibitem{802.11u}
LAN/MAN Standards~Committee of~the IEEE Computer~Society.
\newblock {Wireless LAN Medium Access Control (MAC) and Physical Layer (PHY)
  Specifications, Amendment 9: Interworking with External Networks}.
\newblock IEEE Std 802.11u, February 2011.

\bibitem{PEAPv2}
A.~Palekar, D.~Simon, J.~Salowey, H.~Zhou, G.~Zorn, and S.~Josefsson.
\newblock {Protected EAP Protocol (PEAP) Version 2}.
\newblock Internet draft draft-josefsson-pppext-eap-tls-eap-10, October 2004.
\newblock
  \url{https://tools.ietf.org/html/draft-josefsson-pppext-eap-tls-eap-10}.

\bibitem{Prytuluk2018}
Matt Prytuluk.
\newblock {Two New Security Categories: DNS tunneling VPN and Potentially
  Harmful}.
\newblock Cisco Umbrella, November 2017.
\newblock
  \url{https://support.umbrella.com/hc/en-us/articles/115001077988-Two-New-Security-Categories-DNS-tunneling-VPN-and-Potentially-Harmful}.

\bibitem{rfc8446}
E.~Rescorla.
\newblock {The Transport Layer Security (TLS) Protocol Version 1.3}.
\newblock RFC 8446 (Proposed Standard), August 2018.

\bibitem{rfc2865}
C.~Rigney, S.~Willens, A.~Rubens, and W.~Simpson.
\newblock {Remote Authentication Dial In User Service (RADIUS)}.
\newblock RFC 2865 (Draft Standard), June 2000.

\bibitem{Rivera2017}
Diego Rivera.
\newblock {Are captive portals a security challenge?}
\newblock Intraway, May 2017.
\newblock
  \url{https://thinkincredible.intraway.com/blog-post/captive-portals-security-challenge}.

\bibitem{rfc6648}
P.~Saint-Andre, D.~Crocker, and M.~Nottingham.
\newblock {Deprecating the ``X-'' Prefix and Similar Constructs in Application
  Protocols}.
\newblock RFC 6648 (Best Current Practice), June 2012.

\bibitem{rfc6960}
S.~Santesson, M.~Myers, R.~Ankney, A.~Malpani, S.~Galperin, and C.~Adams.
\newblock {X.509 Internet Public Key Infrastructure Online Certificate Status
  Protocol - OCSP}.
\newblock RFC 6960 (Proposed Standard), June 2013.

\bibitem{rfc4211}
J.~Schaad.
\newblock {Internet X.509 Public Key Infrastructure Certificate Request Message
  Format (CRMF)}.
\newblock RFC 4211 (Proposed Standard), September 2005.

\bibitem{Shamir84}
A.~Shamir.
\newblock {Identity-based cryptosystems and signature schemes}.
\newblock In {\em Advances in Cryptology: CRYPTO '84 (LNCS 196)}, volume~84,
  pages 47--53. Springer, 1984.

\bibitem{rfc5216}
D.~Simon, B.~Aboba, and R.~Hurst.
\newblock {The EAP-TLS Authentication Protocol}.
\newblock RFC 5216 (Proposed Standard), March 2008.

\bibitem{Tan03}
TK. Tan and B.~Bing.
\newblock {Wi-Fi Hotspots}.
\newblock {\em The World Wide Wi-Fi: Technological Trends and Business
  Strategies}, pages 75--95, 2003.

\bibitem{WISPr}
Richard Thurston.
\newblock {WISPr 2.0 boosts roaming between 3G and Wi-Fi}.
\newblock ZDNet, June 2010.
\newblock
  \url{http://www.zdnet.com/article/wispr-2-0-boosts-roaming-between-3g-and-wi-fi}.

\bibitem{Want11}
R.~Want.
\newblock {Near field communication}.
\newblock {\em IEEE Pervasive Computing}, 10(3):4--7, July 2011.

\bibitem{welive2015}
WeLiveSecurity.
\newblock {10 steps to staying secure on public Wi-Fi}, September 2015.
\newblock
  \url{https://www.welivesecurity.com/2015/09/02/10-steps-staying-secure-public-wi-fi}.

\bibitem{httpwg2016}
wififreak.
\newblock {Captive Portals}.
\newblock IETF HTTP Working Group Wiki, February 2016.
\newblock \url{https://github.com/httpwg/wiki/wiki/Captive-Portals}.

\bibitem{WirelessPhreak2017}
WirelessPhreak.
\newblock {Captive Portal and HSTS difficulties}, May 2017.
\newblock
  \url{https://www.wirelessphreak.com/2017/05/captive-portal-and-hsts-issues.html}.

\bibitem{Xia2004}
H.~Xia and J.~Brustoloni.
\newblock {Detecting and Blocking Unauthorized Access in Wi-Fi Networks}.
\newblock In {\em Proc. of the Third IFIP-TC6 Int. Conf. on Research in
  Networking (Networking 2004)}, Athens, Greece, May 2004.

\end{thebibliography}
\bibliographystyle{unsrt}

\end{document}